\documentclass[amsmath,amssymb,floatfix,prl,epsf,12pt]{article}
\usepackage{graphics}
\usepackage{amsmath,amssymb,bm,graphicx,url,epsfig}
\usepackage[active]{srcltx}

\usepackage{bm}

\newcommand{\rf}[1]{(\ref{#1})}
\newcommand{\beq}{\begin{equation}}

\newcommand{\eeq}{\end{equation}}
\newcommand{\bea}{\begin{eqnarray}}
\newcommand{\eea}{\end{eqnarray}}

\newcommand{\e}{\mbox{e}}
\renewcommand{\d}{\mbox{d}}

\newcommand{\Lam}{\Lambda}
\renewcommand{\b}{\beta}
\renewcommand{\a}{\alpha}

\newcommand{\m}{\mu}

\newcommand{\del}{\delta}

\newcommand{\sg}{\sigma}

\newcommand{\oh}{\frac{1}{2}}

\newcommand{\dg}{\dagger}

\newcommand{\tr}{\mathrm{tr}\,}

\newcommand{\no}{\nonumber}
\newcommand{\nn}{\no\\}
\newcommand{\non}{\nonumber \\}

\begin{document}

\begin{center}
\vspace{24pt}
{ \large \bf 
The  XXZ Heisenberg model on  random surfaces. 

}

\begin{center}

\vspace{18pt}

{\sl\bf  J. Ambj\o rn}$\,^{a,c}$, 
and {\sl\bf  A. Sedrakyan}$\,^{a,b}$

\vspace{18pt}
{\footnotesize

$^a$~The Niels Bohr Institute, Copenhagen University\\
Blegdamsvej 17, DK-2100 Copenhagen, Denmark.\\
{ email: ambjorn@nbi.dk}\\

\vspace{10pt}

$^b$~Yerevan Physics Institute\\
Br. Alikhanyan str 2, Yerevan-36, Armenia.\\
{ email: sedrak@nbi.dk}\\

\vspace{10pt}

$^c$~Institute for Mathematics, Astrophysics and Particle Physics (IMAPP)\\ 
Radbaud University Nijmegen \\ 
Heyendaalseweg 135, 6525 AJ, Nijmegen, The Netherlands

}
\end{center}

\vspace{48pt}

\end{center}


\begin{center}
{\bf Abstract}
\end{center}
We consider integrable models, or in general any model  
defined by an $R$-matrix, on random
surfaces, which are discretized using random Manhattan lattices. The set of
random Manhattan lattices is defined as the set dual to the lattice random
surfaces embedded on a regular d-dimensional lattice. They can also be
associated with the random graphs of multiparticle scattering nodes. As an
example we formulate a random matrix model where the partition function
reproduces the annealed average of the XXZ Heisenberg model over all random
Manhattan lattices. A technique is presented which reduces the random matrix
integration in partition function to an integration over their eigenvalues. 
\newpage

\section{Introduction}
\label{intro}

One of the major goals of non-critical string theory
was to describe the non-perturbative physics of
non-Abelian gauge fields in two, three and four dimensions.
The asymptotic freedom of these theories allowed us
to understand the scattering observed at
high energies \cite{Gross, Politzer},
but it also made the long distance, low energy
sector of the theories non-perturbative and indicated
a non-trivial structure of the vacuum \cite{Savvidy,ao}.
It became necessary to develop non-perturbative tools
which would allow us to study phenomena associated with
e.g. confinement. One possibility which attracted a lot of
attention was the attempt by Polyakov to reformulate the
non-Abelian gauge theories as a string theory. This line
of research led Polyakov to his seminal work on non-critical
string theory \cite{Polyakov-1981}. He showed that
the presence of the conformal anomaly forces us  to
include the conformal factor in the string path integral,
and that the action associated with the conformal factor is
the Liouville action. In his approach the study of non-critical
string theory becomes equivalent to the study of two-dimensional
quantum gravity (governed by the Liouville theory) coupled to certain
conformal matter fields.

Attempts to understand and define rigorously the quantum Liouville
theory triggered the lattice formulation, presenting the
two-dimensional random surfaces appearing in the string path
integral as a sum over triangulated piecewise linear surfaces
\cite{Ambjorn, Kazakov,David}. This sum over ``random triangulations''
(or ``dynamical triangulations'' (DT))
could be represented by matrix integrals and in this way
certain matrix integrals became almost synonymous to non-critical
string theory. Somewhat surprisingly many of the lattice models
were exactly solvable and at the same time it was possible
to solve the continuum quantum Liouville model via
conformal bootstrap, and whenever results of the two non-pertubative
methods could be compared, agreement was found.
However, the solutions only made sense  for matter fields
with a central charge $c \leq 1$ \cite{KPZ}. Thus, in a certain way
the approach of Polyakov came to a stalemate with regards
to the question of using non-critical string theory to 
understand the non-perturbative aspects of three 
and four-dimensional non-Abelian gauge theories, which seemingly 
require $c > 1$. It is thus of great interest to try to study
new classes of random surface models which might allow
us to penetrate the $c=1$ barrier. This is one of the main
motivations of this paper. We propose to consider a new class
of random lattices, the so-called random Manhattan lattices.
One is led to such lattices by studying the 
random surface representation of the 3d Ising model  
on a regular 3d lattice \cite{Kavalov-1987, AS-1999},
and via the study of the  Chalker-Coddington network model
\cite{CC-1988}. The study of the latter model led to the
idea that an $R$-matrix could be associated to  a random Manhattan
lattice, and we will consider how to couple in general
a matter system defined by an $R$-matrix to a random lattice.
By summing over the random lattices (i.e. taking the annealed average)
we thus introduce a coupling between the integrable model and two-dimensional
quantum gravity.

\begin{figure}[t]
\centerline{\includegraphics[width=75mm,angle=0,clip]{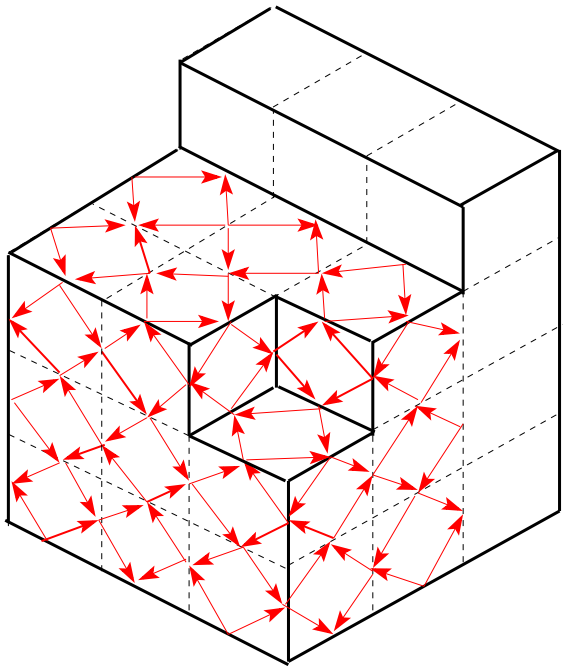}}
\caption{A random surface on the 3d cubic lattice and the 
construction of the dual lattice surface, as described in the text.}
\label{fig1}
\end{figure}

More precisely we start with an integrable model on a 2d square lattice,
assuming we know the R-matrix. We then show that the same R-matrix
can be used on a so-called Random Manhattan Lattice (RML) (see Fig.\
\ref{fig1}), which is a lattice where the links have fixed arrows which
indicate the allowed  fermion hopping.
No hopping is allowed in directions opposite to arrows.
The summation over the RMLs can be performed by a certain matrix integral
related to the R-matrix. This matrix integral is  somewhat different
from the the conventional matrix integrals used to describe conformal
field theories with $c <1$ coupled to 2d quantum gravity, and thus there
is  hope than one can penetrate to $c=1$ barrier. Below we describe the
construction in detail.

\section{The model}

\begin{figure}[h]
\center
\centerline{\includegraphics[width=75mm,angle=0,clip]{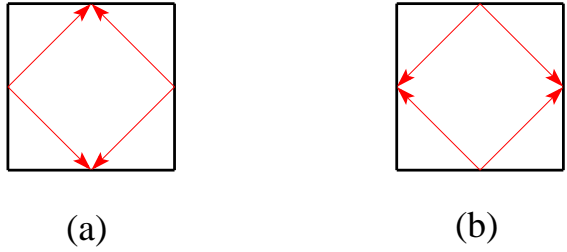}}
\caption{Assignment of arrows to dual lattice}
\label{fig2}
\end{figure}

As mentioned one arrives in a natural way to a RML
from the study of the 3d Ising model on a  regular cubic lattice.
The high temperature expansion of the Ising model can be expressed as a
sum over random lattice surfaces of the kind shown in Fig.\ \ref{fig1},
and on these two-dimensional
lattice surfaces one constructs a kind of dual lattice
by the following procedure:
The lattice surface consists of plaquettes.
Consider the mid-points of the links on the plaquettes as sites of the
dual lattice, and consider arrows on the links as shown in Fig.\ \ref{fig2}.

\begin{figure}[t]
\center
\centerline{\includegraphics[width=75mm,angle=0,clip]{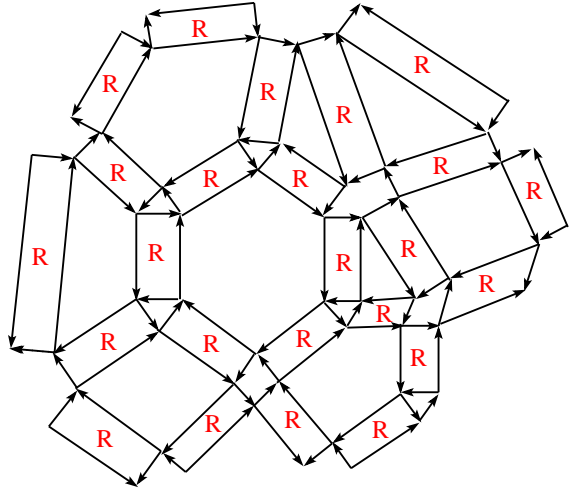}}
\caption{Random Manhattan lattice}
\label{fig3}
\end{figure}

The allover orientation of arrows on the plaquettes should
be such that the flow to neighbouring
plaquettes is continuous as illustrated in Fig.\ \ref{fig3}. This
type of dual lattice with arrows will be a finite Manhattan lattice
corresponding to a particular plaquette lattice
surface on the regular three-dimensional lattice. There is a one to one
correspondence between the plaquette surfaces on the regular lattice and
the finite Manhattan lattices described above.

\begin{figure}[t]
\center
\centerline{\includegraphics[width=95mm,angle=0,clip]{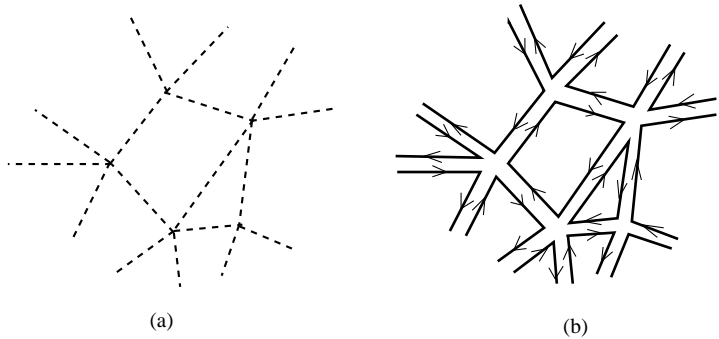}}
\caption{Manhattan lattices and double line graphs}
\label{fig4}
\end{figure}

A second way of obtaining a RML is by starting from
oriented double line graphs, like the ones
introduced by 't Hooft, and then modify the double line propagator like
shown in Fig.\ \ref{fig4}.

\begin{figure}[th]
\center
\centerline{\includegraphics[width=95mm,angle=0,clip]{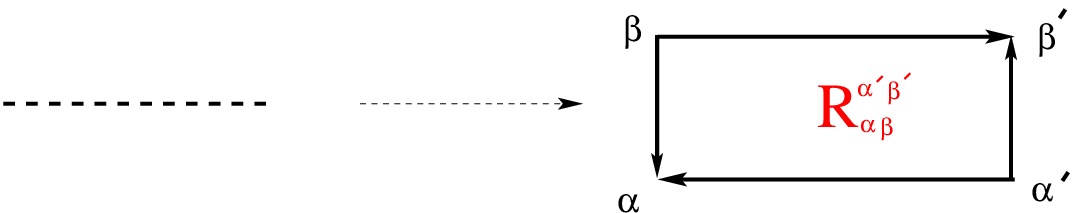}}
\caption{Index assignment of the $R$-matrix}
\label{fig5}
\end{figure}

We will now attach an $R$-matrix of an integrable model to the squares of
the RML with the index assignment shown in Fig.\ \ref{fig5}. Two neighbouring
squares will share one of indices, and the same is thus the case for
the corresponding $R$-matrices, and a summation over values of the indices
are understood, resulting in a matrix-like multiplication of $R$-matrices.
To a RML $\Omega$ we now associate the partition function
\beq\label{1}
Z(\Omega) =  \prod_{R\in \Omega} \check{R},
\eeq
where the summation over indices is dictated by the lattice.
Our final partition function is defined by summing over all
possible (connected) lattices $\Omega$:
\beq\label{2}
Z = \sum_\Omega Z(\Omega) \, \e^{-\mu |\Omega|}
\eeq
where $\mu$ is a ``cosmological'' constant which monitors
the typical size $|\Omega|$ of the lattice $\Omega$. 
As long as we restrict the topology of lattices $\Omega$ entering in the 
sum \rf{2}, there will exist a critical $\mu_c$ such that the sum in \rf{2}
is convergent for $\mu > \mu_c$ and divergent for $\mu < \mu_c$.
We will be interested in a limit where the average value 
of $\Omega$ becomes infinite, and this limit is obtained
by approaching $\mu_c$ from above.
The summation over the elements in $\Omega$, i.e.\ the summation over
a certain set of random 2d lattices, is a regularized version of
the sum over 2d geometries precisely in the same way as in
ordinary DT. It is known the sum over random polygons (triangles, squares,
pentagons etc) with positive weights under quite general conditions
leads to the correct continuum limit, i.e.\ the
functional integral over 2d geometries, when the link length goes to zero.
Thus it is natural to assume that the sum over RML will also
represent in   correct way the sum over 2d geometries when the link length
goes to zero. Under this assumption we have coupled  a given  two-dimensional
model, integrable on a regular lattice, to two-dimensional
quantum gravity. In the next Section we will, in order to make the discussion
more explicit, consider the  {\it XXZ} Heisenberg model, where the R-matrix
is known.

\section{The matrix model}

In order to represent the {\it XXZ} model as a matrix model which
at the same time will offer us a topological expansion of the
surfaces spanned by the oriented ribbon graphs considered above,
we consider the set of $2N\times 2N$ normal matrices. We label the entries
of the matrices as $M_{\a\b,ij}$, where $\a,\b$ takes values 0,1 and
$i,j$ takes values $1,\ldots,N$. The $\a,\b$ indices refer to the {\it XXZ}
model, while the $i,j$ indices will be used to monitor the topological
expansion. A normal matrix is a matrix with complex entries which can
be diagonalized by a unitary transformation, i.e. for a given
normal matrix $M$ there exists a decomposition
\beq\label{3}
M  = U M^{(d)} U^\dg
\eeq
where $U$ is a unitary $2N\times 2N$ matrix and
$M^{(d)}$ a diagonal matrix with eigenvalues
$m_{\a,ii}^{(d)}$ which are complex numbers.

Consider now the action
\beq\label{4}
S(M) = M^*_{\a\b,ij} \check{R}^{\a'\b'}_{\a\b} M_{\b'\a',ij} -
\sum_{n=3}^\infty \frac{d_n}{n}\; \tr \Big(M^n+(M^\dg)^n\Big).
\eeq
We denote the  sum over traces of $M$ and $M^\dg$ as the potential.
The matrix partition function is defined by
\beq\label{4a}
Z = \int \d M \; \e^{-N S(M)}.
\eeq
When one expands the exponential of the potential terms and carries
out the remaining Gaussian integral one will generate all
graphs of the kind discussed above, with the $R$-matrices attached to the
graphs as described. The only difference is that the graphs will be
ordered topologically such that the surfaces associated with the
ribbon graphs appear with a weight $N^\chi$, where $\chi$ is the
Euler characteristics of the surface. If we are only interested in
connected surfaces we should use as the partition function
\beq\label{4b}
F = \log Z.
\eeq
In particular the so-called large $N$
limit, which selects connected surfaces with maximal
$\chi$, will sum over  to
the planar (connected) surfaces  generated by $F$, since these
are the connected surfaces with the largest $\chi$.

Explicitly, in the case of the {\it XXZ} Heisenberg model
the $R$ matrix  is given by:
\bea
\check{R}^{\a'\b'}_{\a\b} &=& \frac{a+c}{2} 1^{\a'}_\a \otimes 1^{\b'}_\b
+\frac{a-c}{2} (\sg_3)^{\a'}_\a \otimes (\sg_3)^{\b'}_\b\nonumber\\
&&\frac{b}{2}\Big((\sg_1)^{\a'}_\a \otimes (\sg_1)^{\b'}_\b
+(\sg_2)^{\a'}_\a \otimes (\sg_2)^{\b'}_\b\Big).
\label{5}\eea

\vspace{12pt}


As an abbreviation we will write
\beq\label{5a}
\check{R}^{\a'\b'}_{\a\b}= \sg_a \otimes \sg_a \tilde{I}_a
\eeq
where a summation over index $a$ is understood, $\sg_0 = 1$, the identity
matrix, and $\sg_a$, $a=1,2,3$ are the Pauli matrices.

Our aim is to decompose the integration over the matrix entries
of $M$ into their radial part $M_d$ and the angular $U$-parameters.
This decomposition is standard, the Jacobian is the so-called Vandermonde
determinant (also in the case of normal matrices, \cite{wiegmann})
When we make that decomposition
the potential will only depend on the eigenvalues $m_{\a,ii}^{(d)}$ and for the
measure we have:
\beq\label{6}
\d M = \d U\;\prod_{\a,i} \d m_{\a,ii}^{(d)} \; \d m_{\a,ii}^{(d)*} \prod_{\a,i \neq \b,j}
\Big|(m_{\a,ii}^{(d)}-m_{\b,jj}^{(d)})\Big|^2.
\eeq
However, the problem compared to a standard matrix integral is that
the matrices $U$, introduced by the transformation \rf{3}, will appear
quartic in the action \rf{4}. Thus the $U$-integration does not reduce to
an independent factor, decoupled from the rest. Neither is it
of the Itzykson-Zuber-Charish-Chandra type.

In order to perform the integral we pass from the transformation
\rf{3} which is given in the fundamental representation, to a
form where we use the adjoint representation. Let us choose a basis
$t^A$ for Lie algebra of the unitary group $U(2N)$ in the fundamental
representation. The normal matrix $M$ can also be expended
in this basis:
\beq\label{6a}
M = C_A t^A,~~~~\tr t^At^B = \del^{AB},
\eeq
where the last condition just is a convenient normalization.
For a given $U$ belonging to the fundamental representation of $U(2N)$
the corresponding matrix in the adjoint representation, $\Lam(U)$,
and the transformation \rf{3} are given by
\beq\label{6b}
\Lam(U)_{AB} = \tr t^A U t^B U^\dg,~~~~C_A = \Lam_{AB} C_B^{(d)},
\eeq
where $C_B^{(d)}$ denotes the coordinates of $M^{(d)}$ in the decomposition
\rf{6a}. The transformation \rf{3} is now
linear in the adjoint matrix $\Lam$ and
the action \rf{4} will be quadratic in $\Lam$. However, we pay of course
a price, namely that the entries of the $(2N)^2\times (2N)^2$
unitary matrix $\Lam$ satisfy more complicated constraints than
those satisfied by the entries of the $2N \times 2N$ unitary matrix $U$.
We will deal with the this problem below. First we express
the action \rf{4} in terms of the eigenvalues
$m_{\a,ii}^{(d)}$ and $\Lam$.

\begin{figure}[th]
\center
\centerline{\includegraphics[width=95mm,angle=0,clip]{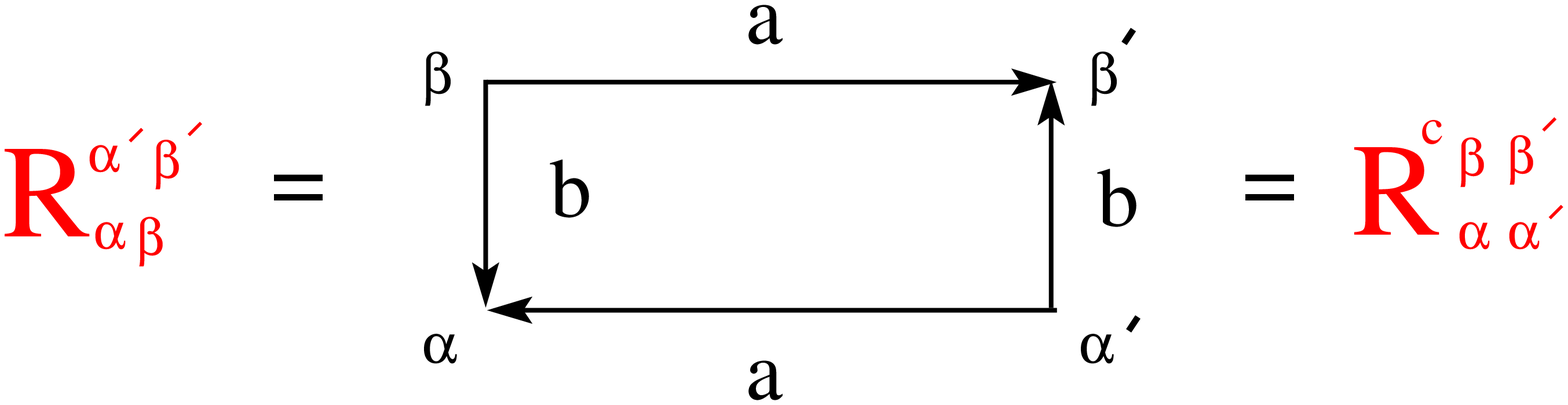}}
\caption{Index assignment of the $R$-matrix}
\label{fig6}
\end{figure}
For this purpose it is convenient to pass from the representation
\rf{5} of the $R$-matrix to the cross channel  (see Fig.\ \ref{fig6}), which
is achieved by:
\beq\label{7}
\left(\check{R}^c\right)^{\b\b'}_{\a\a'} = \check{R}^{\a'\b'}_{\a\b},
\eeq
which amounts to making a particle-hole transformation $0\leftrightarrow 1$
for the indices $\a'$ and $\b$.  After some algebra one obtains:
\bea
\left(\check{R}^c\right)^{\b\b'}_{\a\a'} &=& \frac{b+c}{2}
\sg_1\otimes \sg_1 + \frac{b-c}{2}\sg_2\otimes \sg_2 \nonumber \\
&& \frac{a}{2} \left[ 1\otimes 1 + \sg_3\otimes \sg_3 \right].
\eea
We can now write
\beq\label{9}
\left(\check{R}^c\right)^{\b\b'}_{\a\a'} = \sg_a \otimes \sg_a \; I^a,
\eeq
where the summation is over $a$, $a=0,1,2,3$, $\sg_0 \equiv 1$, the identity
$2\times 2$ matrix. One has:
\beq\label{10}
I_a = \left( \frac{a}{2}, \frac{b+c}{2},\frac{b-c}{2},\frac{a}{2}\right)
\eeq

Using the cross channel $R$-matrix \rf{9} the action \rf{4} becomes:
\beq\label{11}
S= M^*_{\a\b,ij}
\left(\check{R}^c\right)^{\b\b'}_{\a\a'} \del^{i'}_i \del^{j'}_{j}
M_{\b'\a',i'j'} - {\rm Re} V(M).
\eeq

Let $\tau^\m$, $\m=1,\ldots,N^2$ denote generators of the Lie algebra of
$U(N)$, appropriately normalized such that
\beq\label{12}
\del^{i'}_{i} \del^{j'}_{j} = \tau^\m_{ij}\tau^\m_{i'j'}.
\eeq
If we insert \rf{12} into the action \rf{11} we obtain
\beq\label{13}
S= \tr \left( M^\dg\sg^a \tau^\m\right) I^a
\tr \left(\sg^a \tau^\m M\right)
\eeq
where we can view $\sg^a\tau^\m$, $a=0,1,2,3$ and $\m=1,\ldots,N^2$ as
the $(2N)^2$ generators $t^A$ of the Lie algebra of $U(2N)$.
Formulas \rf{6a} and \rf{6b} with the generators
$\sg^a\tau^\m$ read
\beq\label{15}
m^{(d)}_{a,\m} = \oh \tr \left(M^{(d)} \sg^a \tau^\m \right),
\eeq
\beq\label{16}
\Lam^{a\m,a'\m'} =
\oh \tr \left(\sg^a \tau^\m U  \sg^{a'} \tau^{\m'} U^\dg \right).
\eeq
We now want to use \rf{3} and \rf{15} and \rf{16} to express
the matrix $M$ in the action \rf{13} in terms of $m^{(d)}_{a,\m}$ and
$\Lam^{a\m,a'\m'}$, and we obtain
\beq
S= m^{(d) *}_{a'\m'} \Lam^{a\m,a'\m'} I^a \Lam^{a\m,a''\m''}m^{(d)}_{a''\m''}
-V(m^{(d)}_{a\m})
\label{14}
\eeq

It is convenient to choose following basis for generators $ t_A,\; A \equiv (a, ij)=1\cdots 4 N^2$ 
of $ U(2 N)$. For
Cartan sub-algebra we take $t_{0,11}=1\otimes 1$ for common phase factor,  $t_{3,ii}=\sg_3 \otimes \tau^{ii} ,\; i=1\cdots N$
and $t_{0,ii}=\frac{1}{2}(1+\sg_3)\otimes (\tau^{i,i}-\tau^{i-1,i-1}), \; i=2\cdots N $ for other $ 2 N-1$ traceless generators,
where
\beq\label{17}
\left(\tau^{ij}\right)_{kl} = \del_{ik}\del_{jl}.
\eeq
For the remaining, non-diagonal generators we take $\sg_a \tau^{ij} ,\; a=1,2,\; i,j=1\cdots N$ and
 $\sg_a \tau^{ij} ,\; a=0,1,2,3, \; i \neq j=1\cdots N$.
Then, for this choice of generators and from 
$m^{(d)}_{a,ij} = \oh \tr \left(M^{(d)} t_{a,ij}\right) $
 we have
\bea\label{17a}
m^{(d)}_{0,11} &=& \tr M=\sum (m^{(d)}_{\a=1,ii}+m^{(d)}_{\a=2,ii})\\
m^{(d)}_{3,ii} &=& m^{(d)}_{\a=1,ii}-m^{(d)}_{\a=2,ii},\\
m^{(d)}_{0,ii} &=& m^{(d)}_{\a=1,ii}-m^{(d)}_{\a=1,i-1 i-1} \label{17b}\\
m^{(d)}_{b=1,ii} &=& m^{(d)}_{b=2,ii} = 0\label{17c}
\eea
with the rest of elements $m^{(d)}_{a,ij}=0,\; a=0,1,2,3,\; i\neq j=1,\cdots N$. 

\subsection{The space of integration}

We now change the integration over the unitary matrices $U(2N)$ in
formula \rf{6}, which are in the fundamental representation, to the
unitary matrices $\Lam(2N)$ in the adjoined representation.
Rather than using the Haar measure expressed in terms of the $U$-matrices
we should express the Haar measure in terms of the $\Lam$-matrices.

Since normal matrices $M$ can be regarded as  elements in the algebra $u(2N)$,
the action of $\Lam$, defined by the formula (\ref{6b}) on its
diagonalized form (\ref{17a}), will form an orbit
in the algebra with the basis consisting of all diagonal matrices.
Diagonalized elements of $M$ are invariant under the action of the maximal
abelian (Cartan) subgroup $\otimes U(1)^{2N}$of $U(2N)$.
Therefore the orbits are isomorphic to the factor space
$\frac{U(2N)}{U(1)\,\otimes \cdots \otimes \,U(1)}$.

Moreover this factor space is isomorphic
to a so-called flag-manifold, defined as follows
(see \cite{Brion, Fuks} and references there):
A single flag is a sequence of nested complex subspaces in a
complex vector space $C_n$
\bea
\label{flag}
\{\varnothing\}=C_0 \subset C_{a_1} \subset \cdots \subset C_{a_k}
\subset C_n=C^n
\eea
with complex dimensions $dim_C C_i=i$.
For a fixed set of integers $(a_1, a_2 \cdots a_k, n)$ the collection of all
flags forms a manifold, which called a flag manifold $F(n_1,n_2,\cdots n_k)$,
where  $n_i=a_i-a_{i-1} $.
The manifold $F(1,1,\cdots 1)$ is called a full flag manifold,
others are partial flag manifolds.
The full flag manifold $F(1,1,\cdots 1)$ is isomorphic
to the orbits of the action of the adjoined
representation of $U(2 N)$ on its algebra
\bea
\label{orbit}
F(1,1,\cdots 1) = \frac{U(2 N)}{\otimes U(1)^{2N}}
\eea

The set of $C_{i}$ hyperplanes in $C_{i+1}$ is isomorphic
to the set of complex lines in $C_{i+1}$.
In differential geometry this set is denoted by  $\mathbf{CP}^{i}$ (and
also  as Grassmanians $\mathbf{Gr}(1,i)$) and is called a
complex projective space.  Hence, the complex projective
space is a factor space
\bea
\label{cp}
\mathbf{CP}^{i}= \frac{U(i+1)}{U(i)\otimes U(1)}= \frac{S^{2 i+1}}{U(1)}
\eea
where $S^{2 i+1}$ is a real $2 i+1$ dimensional sphere.

According to description presented above the orbit of
the action of the adjoined representation
of  $U(2 N)$ on the set of normal matrices $M$ is a sequence of
fiber bundles and locally, on suitable
open sets,  the elements of the flag manifold
can be represented as a direct product of projective spaces
(the fibers)
\bea
\label{FB}
\frac{U(2 N)}{\otimes \,U(1)^{2N}} \simeq \mathbf{CP}^{2 N-1} \times \mathbf{CP}^{2 N-2}
\times \cdots \mathbf{CP}^{1}
\eea

In simple words we have the following representation of the orbit:
any diagonalized normal matrix in the
adjoined representation has a following form
\bea
\label{diagonal-form}
M^{(d)}_{a\mu}&=&\Big(\underbrace{m^{(d)}_{3,NN},0,\cdots 0}_{4 N-1},\underbrace{m^{(d)}_{0,NN},0,\cdots 0}_{4 N-3},
\cdots \underbrace{m^{(d)}_{3,kk},0,\cdots 0}_{4 k-1},\underbrace{m^{(d)}_{0,kk},0,\cdots 0}_{4 k-3},\non
 &\cdots&\underbrace{m^{(d)}_{3,11},0,0}_{3},m^{(d)}_{0,11}\Big)
\eea
The action of the adjoined representation $\Lam$ on this
$M$ transforms it into the elements of $\frac{U(2 N)}{\otimes U(1)^{2N}}$
presented in (\ref{FB}) where $\mathbf{CP}^{2 k-1}$
represents image of the part $\underbrace{m^{(d)}_{3,kk},0,\cdots 0}_{4 k-1} $.

This implies that the measure of our integral
over normal matrices $M$ can be decomposed
into the product of measures of the base space (the diagonal matrices)
and the flag manifold (the fiber)
\bea
\label{measure}
{\cal D}\Lam &=& \prod_{i=1}^{N} d m^{(d)}_{0,ii}d m^{(d)}_{3,ii} \prod_{k=1}^{2 N-1} {\cal D}[\mathbf{CP}^{k}]\non
&=&\prod_{i=1}^{N} d m^{(d)}_{0,ii}d m^{(d)}_{3,ii} \prod_{k=1}^{N} {\cal D}\Big[\frac{S^{4 k-1}}{S^1}\Big]{\cal D}\Big[\frac{S^{4 k-3}}{S^1}\Big]
\eea

However, since the diagonal matrix elements
$m^{(d)}_{0,ii} $ and $m^{(d)}_{3,ii} $ are complex,
our action is invariant over $\otimes \, U(1)^{2N}$ (one $U(1)$
per marked segment in (\ref{diagonal-form})
and we can extend the integration measure from (\ref{measure}) to
\bea
\label{measure-2}
{\cal D}\Lam =\prod_{i=1}^{N} d m^{(d)}_{0,ii}d m^{(d)}_{3,ii}
\prod_{k=1}^{N} {\cal D}\big[S^{4 k-1}\Big]{\cal D}\big[S^{4 k-3}\Big]
\eea
In other words, we suggest that the action of $\Lam$
on the segments $\underbrace{m^{(d)}_{3,kk},0,\cdots 0}_{4 k-1}$
and $\underbrace{m^{(d)}_{0,kk},0,\cdots 0}_{4 k-3}$ in
(\ref{diagonal-form}) form vectors
$m^{(d)}_{3,kk} z_{3,k}^r,\; (r=1 \cdots 4 k-1) $
and $m^{(d)}_{0,kk} z_{0,k}^s,\; (s=1 \cdots 4 k-3) $, respectively,
where the real coordinates $z_{3,k}^r$ and $z_{0,k}^r$ belong 
to the unite spheres $S^{4k-1}$ and $S^{4k-3}$.

In order to write the measure of integration over the
spheres $S^{4k-1}$ and $S^{4k-3}$ we embed them into the Euclidean spaces
$R^{4k}$ and $R^{4k-2}$, respectively, and define
\bea
\label{measure-3}
&&{\cal D}\big[S^{4 k-1}\Big] =
\delta\Big(\sum_{s=1}^{4 k} [z_{3,k}^s]^2-1\Big)\prod_{s=1}^{4 k} dz_{3,k}^s
=\int d\lambda_{3,k}\prod_{s=1}^{4 k-1} dz_{3,k}^s 
e^{-\lambda^2_{3,k}(1-\sum_{s=1}^{4 k} [z_{3,k}^s]^2)},\nonumber \\
&&{\cal D}\big[S^{4 k-3}\Big] =
\delta\Big(\sum_{s=1}^{4 k-2} [z_{0,k}^s]^2-1\Big) 
\prod_{s=1}^{4 k-2} dz_{0,k}^s
=\int d\lambda_{0,k}\prod_{s=1}^{4 k-2} dz_{0,k}^s 
e^{-\lambda^2_{0,k}(1-\sum_{s=1}^{4 k-3} [z_{0,k}^s]^2)}, \nonumber\\
\eea
where we have introduced Gaussian integrations over the 
real parameters $\lambda_{a,k},\; a=0,3 $.  These integrations 
reproduce the  factors 
$ \frac{1}{2 \sqrt{1-\sum_{s=1}^{4 k-3} [z_{a,k}^s]^2}}$ which arise from the 
$\delta$-functions in \rf{measure-3} by  integrations over the  coordinates 
$ z_{3,k}^{4k}$ and $ z_{0,k}^{4k-2}$. 
We  have omitted coefficients
$\sqrt{\pi}/2$ in front of integrals on the  right hand side of 
the expressions (\ref{measure-3}) since they
unimportant for the partition function.

With this definition of the measure the partition function (\ref{4a})
can be written as
\bea\label{21b}
\int \d M e^{-N S(M)} &=&
\int \prod_{k,a=0,3} \d m^{(d)}_{a,kk} \d\lambda_{a,k} \prod_{s=1}^{4 k-1} dz_{3,k}^s
\prod_{s=1}^{4 k-3} dz_{0,k}^s W(m^{(d)}_{\alpha,kk}) e^{-S(m_{a,kk}^{(d)},\lambda_{a,k},z_{a,k})},\non
\eea
where $W(m^{(d)}_{\alpha,kk}) =\prod_{\a,i \neq \b,j}
\Big|(m_{\a,ii}^{(d)}-m_{\b,jj}^{(d)})\Big|^2 $ is Vandermonde determinant and
\bea
\label{action-2}
S(m_{a,kk}^{(d)},\lambda_{a,k},z_{a,k})=
\sum_{a=0,3,k=1}^N \Big[|m^{(d) *}_{a,kk}|^2 
\sum_{b=1}^{s(a,k)}|z_{a,k}^{b}|^2 I^b
+\lambda^2_{a,k}\big(1-\sum_{s=1}^{s(a,k)} [z_{a,k}^s]^2\big)- 
V(m^{(d)}_{a,kk})\Big].\nn
\eea
Here $s(3,k)=4k-1$ and $s(0,k)=4k-3$, while according to (\ref{10})
\bea
\label{Ia}
I=\frac{1}{2}\Big(\underbrace{a, b-c,b+c,a \cdots b+c }_{4 N-1},
\underbrace{a,a,b-c,b+c,a \cdots a}_{4 N-3},
\cdots \underbrace{a,b-a,b+a}_{3},\underbrace{a}_1\Big).
\eea
The length of $I$ precisely is $4 N^2$.
The  4 elements   $a/2,(b-c)/2,(b+c)/2,a/2$ of $I^b$ 
are placed as in a specific sequence in $I$, as shown in eq.\ (\ref{Ia}).
However, the partition function is independent of this choice (which 
is just our choice arbitrary choice) after integration.

As one can see we have in the partition function (\ref{action-2})
simple Gaussian integrals over $z_{a,k}^s$. These can be evaluated
and we are left with integrals over $m^{(d)}_{a,kk}$ and $\lambda_{a,k}$ only.
It is convenient to rescale the Lagrange multipliers
and introduce  $\tilde{\lambda}_{a,k}= |m^{(d)}_{b,ii}|^{-1}\lambda_{a,k} $.
Then, after performing the Gaussian integrals, we obtain
\bea
\label{27}
Z &=& \int   \prod_{a=0,3;k=1}^N \d m^{(d)}_{a,kk} W(m^{(d)}_{a,kk}) \prod_{k=1}^N \frac{1}{|m^{(d)}_{3,kk}|^{4k-2}|m^{(d)}_{0,kk}|^{4k-4}}
\int \prod_{a=0,3;k=1}^N\d\tilde{\lambda}_{a,k}\; \nonumber\\
&\cdot&\prod_{a=0,3;k=1}^Ne^{-{\rm Re} V(m^{(d)}_{a,kk})- |m^{(d)}_{a,kk}|^2 \tilde{\lambda}^2_{a,k}}
 \prod_{a=0,3;k=1}^N Z_{a,k} (\tilde{\lambda}_{a,k}).
\eea
where
\bea
\label{Z}
Z_{3,k} (\tilde{\lambda}_{3,k})=
\frac{1}{(a- \tilde{\lambda}^2_{3,k} )^{k-\frac{1}{2}}}
\frac{1}{(b-c- \tilde{\lambda}^2_{3,k} )^{\frac{k}{2}}}
\frac{1}{(b+c- \tilde{\lambda}^2_{3,k} )^{\frac{k}{2}}}\nonumber\\
Z_{0,k} (\tilde{\lambda}_{0,k})=
\frac{1}{(a-\tilde{\lambda}^2_{0,k} )^{k-\frac{1}{2}}}
\frac{1}{(b-c- \tilde{\lambda}^2_{0,k} )^{\frac{k-1}{2}}}
\frac{1}{(b+c-\tilde{\lambda}^2_{0,k} )^{\frac{k-1}{2}}}
\eea

Let us demonstrate that the Gaussian integration  
over the adjoint representation matrice $\Lambda$
in (\ref{27}) correctly reproduces the partition function (\ref{4a}) 
when interaction is absent, i.e.\ $V(M)=0$. In this case the integral
over normal matrices $M$ in the fundamental representation of 
$U(2 N)$ can  easily be performed directly  and the result is:
$$  
\frac{1}{\sqrt{\prod_{b=1}^{4 N^2} I_b}}=\big[a^2 (b^2-c^2)\big]^{-N^2/2}.
$$ 

Let us first consider the simple case $N=1$.
In the general setup this corresponds to having the two shortest, 
length 3 and 1 segments in the sequence   (\ref{diagonal-form}).
The Vandermonde determinant cancels the $m$'s 
in the denominator in the expression   (\ref{27}) of the partition function 
and integration over the $m$'s leads to 
\bea
\label{N=1}
Z&=& \int \frac{\d\tilde{\lambda}_{3,1}}{\tilde{\lambda}_{3,1}} 
\frac{\d\tilde{\lambda}_{0,1}}{\tilde{\lambda}_{0,1}}
\frac{1}{(a- \tilde{\lambda}^2_{3,1} )^{1/2}}
\frac{1}{(b-c- \tilde{\lambda}^2_{3,1} )^{1/2}}
\frac{1}{(b+c- \tilde{\lambda}^2_{3,1} )^{1/2}}
\frac{1}{(a- \tilde{\lambda}^2_{0,1} )^{1/2}}\nn
&=& \frac{1}{[a^2(b^2-c^2)]^{1/2}}
\eea
provided we place the $\lambda$-poles at zero and the $\lambda$-branch 
cuts at different sides of the real $\lambda$-axis.

Now consider a general $N$. For a  generic $i$ segment in 
(\ref{diagonal-form}) we first represent the multipliers 
$|m^{(d)}_{1,ii}-m^{(d)}_{1,jj}|$ and  $|m^{(d)}_{1,ii}-m^{(d)}_{2,jj}|$ in the  
Vandermonde determinant as a sum over $m^{(d)}_{a,ii},\; a=0,3 $:   
\bea
\label{last1}
|m^{(d)}_{1,ii}-m^{(d)}_{1,jj}|&=&
|m^{(d)}_{0,ii}+\sum_{r=j+1}^{i-1} m^{(d)}_{0,rr}| \nn 
|m^{(d)}_{\a=1,ii}-m^{(d)}_{\a=2,jj}|&=&
|m^{(d)}_{3,ii}+
\sum_{r=j+1}^{i-1}( m^{(d)}_{\a=2,rr}- m^{(d)}_{\a=2,r-1 r-1})| .
\eea 
When using this decomposition in the 
 Vandermonde determinant  product, we observe that only the  
contribution of the selected terms $ |m^{(d)}_{0,ii}|$ and $ |m^{(d)}_{3,jj}|$
in (\ref{last1}) will cancel all  $m^{(d)}$'s in the denominator of 
(\ref{27}) and thus lead to  a nonzero contribution. By  Cauchy
integration as above we obtain
\bea
\label{N}
Z&=& \int \prod_{a=0,3;i=1}^N  
\frac{\d\tilde{\lambda}_{3,i}}{\tilde{\lambda}_{3,i}} 
\frac{\d\tilde{\lambda}_{0,i}}{\tilde{\lambda}_{0,i}}
 Z_{a,i} (\tilde{\lambda}_{a,i})\nn
&=& \frac{1}{[a^2(b^2-c^2)]^{N^2/2}}.
\eea  
Other terms in the decomposition (\ref{last1}) will result 
in $\tilde{\lambda}_{a,i},\; a=0,3$ appearing in the denominator
of the integral (\ref{N}) with other powers than one, and 
the integration will give zero for these terms.

\section{Conclusions}
  
We have defined a matrix model which reproduces the partition function of an
integrable model (the XXZ model) on  random surfaces.
The random surfaces under consideration appear
as random Manhattan lattices, which are dual to random surfaces
embedded in  a $d$ dimensional
regular Euclidean lattice. This formulation allows us to consider a
new type of non-critical strings with $c >1$.

We have shown that  the matrix integral can be reduced to
integrals over the eigenvalues of matrices.
The important ingredient in the integration
over angular parameters of the matrices
(which are usually defined by the Itzykson-Zuber integral)
is the reduction of the problem to an integration 
over unitary matrices in the adjoined representation.
In principle one can now try to apply standard large 
N saddle point methods for solving the resulting integrals 
over eigenvalues.

Thoughout  in this  paper we have considered the Heisenberg XXZ model 
on random surfaces, but the approach can be applied to other integrable models.


\section{Acknowledgement}
A.S thanks Niels Bohr Institute and Armenian Research Council for 
partial financial support.
JA acknowledges support from the ERC-Advance grant 291092,``Exploring the Quantum Universe'' (EQU). The authors acknowledge support 
of FNU, the Free Danish Research Council, from the grant 
``quantum gravity and the role of black holes''. Finally this research 
was supported in part by the Perimeter Institute of Theoretical Physics.
Research at Perimeter Institute is supported by the Government of Canada
through Industry Canada and by the Province of Ontario through the 
Ministry of Economic Development \& Innovation.


\begin{thebibliography}{99}

\bibitem{Gross} D. Gross, F. Wilczek,  Phys. Rev. Lett. {\bf 30}, 1343 (1973).

\bibitem{Politzer} H. D. Politzer, Phys. Rev. Lett. {\bf 30}, 1346 (1973).

\bibitem{Savvidy}  G. Savvidy, Phys.lett. B {\bf 71},133 (1977),\\
S. Matinyan, G. Savvidy,  Nucl. Phys. B {\bf 134}, 539 (1978).

\bibitem{ao}
  J.~Ambjorn and P.~Olesen,
  Nucl.\ Phys.\ B {\bf 170} (1980) 60;
  Nucl.\ Phys.\ B {\bf 170} (1980) 265.




\bibitem{Polyakov-1981} A. Polyakov, Phys. Lett. B. {\bf 103}, 207 (1981);
Phys. Lett. B. {\bf 103}, 211 (1981).

\bibitem{Ambjorn} J. Ambj\o{}rn, B. Durhuus and J. Fro¨lich, 
Nucl. Phys. {\bf B257}[FS14],433 (1985).

\bibitem{Kazakov} V.A. Kazakov, Phys. Lett. B {\bf 150},282 (1985).

\bibitem{Kazakov-2} V.A. Kazakov, I.K. Kostov and A.A. Migdal, 
Phys. Lett. B {\bf 157},  295 (1985).

\bibitem{David}  F. David, Nucl. Phys. {\bf B257} [FS14],45 (1985).

\bibitem{Migdal} D. Gross and A.A. Migdal, Phys. Rev. Lett. 
{\bf 64},127 (1990);\\
M. Douglas and S. Shenker, Nucl. Phys. B {\bf 335}, 635 (1990).

\bibitem{Kazakov-3} E. Bre´zin and V.A. Kazakov, Phys. Lett.B {\bf 236}, 144 (1990).

\bibitem{KPZ} D. Knizhnik, A. Polyakov, A. Zamolodchikov,   Mod. Phys. Lett. A {\bf 3}, 819 (1988).


\bibitem{Kavalov-1987} A. Kavalov,  A. Sedrakyan, Nucl. Phys. B {\bf 285}, 264 (1987).

\bibitem{AS-1999} A. Sedrakyan - Nucl.Phys. {\bf B 554 [FS]},514 (1999).

\bibitem{Sedrakyan-Como} A. Sedrakyan, Phys. Rev. B {\bf 68}, 235329 (2003),\\
A. Sedrakyan, In Proceedings
 of Advanced NATO Workshop on Statistical
Field Theories, edited by A. Capelli and G. Mussardo, Kluwer
Academic, Amsterdam, 2002.

\bibitem{CC-1988} J. Chalker and P. Coddington, J. Phys. C 21, 2665 (1988).

\bibitem{wiegmann}  P. Wiegmann, A. Zabrodin,  J. Phys. A {\bf 36} 3411 (2003).

\bibitem{IZ} C. Itzykson, J. B. Zuber, J. Math. Phys. {\bf 21}, 411 (1980).


\bibitem{Brion} M. Brion : Lectures on the geometry of flag varieties, In "Topics in Cohomological Studies of Algebraic Varieties",
Trends in Math., Birkhäuser, 2005.

\bibitem{Fuks} D.B. Fuks, In "Itogi Nauki i tekhniki. Serie Sovremennie problemi mathematiki. Fundamentalnie Napravlenia", Vol.12, p.253-314(1986) 
(Moscow, 1986).

\end{thebibliography}
\end{document}